
\input harvmac
\input epsf		
\input tables		

\def\eV {\ifmmode \,\, {\rm eV} \else eV \fi}
\def\keV{\ifmmode \,\, {\rm keV} \else keV \fi}
\def\MeV{\ifmmode \,\, {\rm MeV} \else MeV \fi}
\def\GeV{\ifmmode \,\, {\rm GeV} \else GeV \fi}
\def\TeV{\ifmmode \,\, {\rm TeV} \else TeV \fi}
\def\myinstitution{\vskip 17pt
   \centerline{\it Enrico Fermi Institute and Department of Physics }
   \centerline{\it University of Chicago, 5640 S. Ellis Ave. }
   \centerline{\it Chicago, IL 60637}
   \vskip .4in
}
\def\myemail{\footnote{$^\dagger$}
	{E-mail: jungman@yukawa.uchicago.edu}}

\def\ifig#1{\noindent\Fig#1 \hbox{: } {\reflabeL{ #1}}}
\def\Fig{Fig.~\the\figno\nFig}
\def\nFig#1{\xdef#1{Fig.~\the\figno}%
\writedef{#1\leftbracket Fig.\noexpand~\the\figno}%
\ifnum\figno=1\immediate\openout\ffile=figs.tmp\fi\chardef\wfile=\ffile%
\immediate\write\ffile{\noexpand\medskip\noexpand\item{Fig.\ \the\figno. }
\reflabeL{#1\hskip.55in}\pctsign}\global\advance\figno by1\findarg}



\def\sbar#1{\kern 0.8pt
        \overline{\kern -0.8pt #1 \kern -0.8pt}
        \kern 0.8pt}  
\def\Dx{{d\over dx}}



\lref\cowsik{ H.~Harari and Y.~Nir, Nucl.~Phys. {\bf B292}, 251 (1987)\semi
       R.~Cowsik and J.~McClelland, Phys.~Rev.~Lett. {\bf 29}, 669 (1972).}
\lref\numrec{ W.H.~Press et.al., {\sl Numerical Recipes: The Art of
	Scientific Computing}, Cambridge (1989).}
\lref\obsB{ E.W.~Kolb and M.S.~Turner, {\sl The Early Universe}, Addison-Wesley
(1989),
	and references therein.}
\lref\naive{ This argument is actually more naive than alluded to in the text.
This
	simple behaviour of the asymmetry under scaling of the right-handed neutrino
	masses can break down under extreme conditions, such as are illustrated
	by the non-minimal model which we discuss below; in that case one
	$N_R$ plays a much different role than the others and the basic picture
	changes. }
\lref\solaranal{V.~Barger, R.~J.~N. Phillips, and K.~Whisnant, Phys.~Rev. {\bf
D43}, 1110 (1991). }
\lref\entropyprod{D.~A.~Dicus, E.~W.~Kolb, V.~L.~Teplitz, and R.~V.~Wagoner,
Phys.~Rev. {\bf D17}, 1529 (1978).}
\lref\arnmc{P.~Arnold and L.~McClerran, Phys.~Rev. {\bf D36}, 581 (1987).}
\lref\Peccei{R.~D.~Peccei, UCLA preprint UCLA/92/TEP/33, to appear in the
Proceedings of
		ICHEP XXVI, Dallas, Texas (1992).}
\lref\jarl{ C.~Jarlskog, Phys.~Rev.~Lett., {\bf 55}, 1039 (1985).}
\lref\gassleut{J.~Gasser and H.~Leutwyler, Phys.~Rep. {\bf 87}, 77 (1982).}
\lref\monann{G.~Lazarides and Q.~Shafi, Phys.~Lett. {\bf 94B}, 149 (1980).}
\lref\preskill{ J.~Preskill, Ann.~Rev.~Nucl.~Part.~Sci. {\bf 34}, 461 (1984).}
\lref\patisalam{J.C.~Pati and A.~Salam, Phys.~Rev. {\bf D10}, 275 (1974).}
\lref\frit{H.~Fritzsch, Phys.~Lett. {\bf B73}, 317 (1978).}
\lref\babshaf{K.S.~Babu and Q.~Shafi, Bartol preprint BA-92-27 (1992).}
\lref\GJ{ H.~Georgi and C.~Jarlskog, Phys.~Lett. {\bf B89}, 297 (1974).}
\lref\moaxion{R.N.~Mohapatra and G.~Senjanovic, Z.~Phys. {\bf C17}, 53 (1983).}
\lref\PQ{R.D.~Peccei and H.~Quinn, Phys.~Rev.~Lett. {\bf 38}, 1440 (1977)\semi
	R.D.~Peccei and H.~Quinn,  Phys.~Rev. {\bf D16}, 1791 (1977).}
\lref\nextpaper{ T. Gherghetta and G. Jungman, in preparation.}
\lref\Dpar{ T.W.B.~Kibble, G.~Lazarides, and Q.~Shafi, Phys.~Rev. {\bf D26},
435 (1982)\semi
	    D.~Chang, R.N.~Mohapatra, M.K.~Parida, Phys.~Rev.~Lett. {\bf 52}, 1072
(1984)\semi
	    D.~Chang, R.N.~Mohapatra, M.K.~Parida, Phys.~Rev. {\bf 30}, 1052 (1984).}
\lref\GUTdeltaB{E.W.~Kolb and S.~Wolfram, Nucl.~Phys {\bf B172}, 224
(1980)\semi
		J.A.~Harvey, E.W.~Kolb, D.B.~Reiss, and S.~Wolfram, Nucl.~Phys {\bf B201},
161 (1982).}
\lref\harvturn{ J.~Harvey and M.~Turner, Phys.~Rev. {\bf D42}, 3344 (1990).}
\lref\leptontrouble{M.~Fukujita and T.~Yanagida Phys.~Rev.
	{\bf D42}, 1285 (1990)\semi
	S.~Barr and A.~Nelson, Phys.~Lett. {\bf 246B}, 141 (1990).
}
\lref\markus{ M.A.~Luty, Phys.~Rev. {\bf D45}, 455 (1992).}
\lref\sphalerons{V.~Kuzmin, V.~Rubakov, and M.~Shaposhnikov, Phys.~Lett. {\bf
155B}, 36 (1985)\semi
		F.~Klinkhamer and N.~Manton, Phys.~Rev. {\bf D30}, 2212 (1984).}
\lref\sotenme{G.~Jungman, Phys.~Rev. {\bf D46}, 4004 (1992). }
\lref\babmoh{K.S.~Babu and R.N.~Mohapatra, Bartol preprint BA-92-054 (1992).}

\Title{\vbox{\hbox{ EFI-93-07}\hbox{ hep-ph/9302212}}}{\vbox{
	\centerline{\hbox{Cosmological Consequences of
		Spontaneous}}\vskip 5pt
        \centerline{\hbox{Lepton Number Violation in
		$SO(10)$ Grand Unification}} } }
\def\hisemail{\footnote{$^*$}
        {E-mail: tonyg@yukawa.uchicago.edu}}
\centerline{ Tony Gherghetta\hisemail {\hskip 3pt} {\it and}
	{\hskip 3pt}Gerard Jungman\myemail}
\myinstitution

\noindent
Cosmological constraints on grand unified theories with spontaneous
lepton number violation are analysed. We concentrate on $SO(10)$,
the simplest of the models possessing this property.
It has been noted previously that the consistency of these models
with the observed baryon asymmetry generically implies strict
upper bounds on the light neutrino masses. In this paper, we analyze
the situation in detail. We find that minimal models of fermion
masses face difficulties, but that it is possible for these models
to generate an adequate
baryon asymmetry via non-equilibrium lepton number violating processes
when the right-handed neutrino masses are near their maximum possible
values. This condition uniquely picks out the minimal gauge symmetry breaking
scheme. A non-minimal model is also analyzed, with somewhat different
conclusions due to the nature of the imposed symmetries.

\Date{January 1993}


\newsec{Introduction}

The ability of grand-unified theories to produce the observed
baryon asymmetry of the universe is arguably one of their most
attractive features \GUTdeltaB.
However, for some time it has been realized \leptontrouble\harvturn\
that grand-unified theories which
also exhibit spontaneous lepton number violation are in danger of
eliminating this baryon asymmetry at later times, when one considers
anomalous $B+L$ violating processes \sphalerons\ to be in thermal equilibrium.
Generally this danger is averted by prescribing a bound on the
neutrino masses such that lepton number violating processes are not
in equilibrium at temperatures near or below the scale of the
spontaneous lepton number violation.
The actual situation is somewhat more complicated because such out of
equilibrium processes can create an asymmetry, approximately
proportional to CP violating parameters. A full calculation must be considered.

The purpose of this paper is to
analyze leptogenesis in these models
and the consequent constraints on grand unification,
specifically $SO(10)$ models.
$SO(10)$ is the smallest candidate possessing the
required spontaneous lepton number violation, and many larger groups
can have symmetry breaking chains which contain an $SO(10)$ stage.
We consider the phenomenologically viable breaking pattern
$SO(10) \rightarrow SU(2)_L\times SU(2)_R \times SU(4)$ \patisalam.
$SU(4)$ breaks to $SU(3)_c\times U(1)_{B-L}$,
$SU(2)_R$ breaks to $U(1)_{I_{3R}}$, and the subsequent
breaking of $U(1)_{B-L}\times U(1)_{I_{3R}}$ to $U(1)_Y$
provides the spontaneous violation of lepton number.
Constraints on the intermediate scales arising from gauge
coupling unification and Yukawa coupling unification are
analysed in Ref. \sotenme.

Of course, one must realize that it is hard to make
definitive statements
about the grand-unification program based on fermion mass phenomenology.
Many constraints can be avoided by introducing arbitrary complications
in the Higgs sector.
However, if it is found that many (or all) of the minimal prescriptions
are not viable, then the model can lose its attractiveness.
Therefore, our philosophy is to analyze certain specific prescriptions
for fermion masses, bringing to bear all the known constraints.
After such an analysis we can attempt to abstract those features
which will remain true generally and those which seem specifically
model dependent. It is not too difficult to construct attractive
models which are consistent with all known low-energy parameters,
but it is more difficult to construct models which survive
the various cosmological constraints.

In this paper we consider certain minimal models which
can give rise to realistic fermion masses and mixings.
As discussed in Ref. \sotenme,
a detailed analysis of the situation, with emphasis on neutrino masses and
lepton number violating processes in the early universe,
has been required for some time.
For any such model, there are two issues to address. First, there is the
question of whether
or not the lepton sector mixing implied by a given fermion mass model
is sufficient to render the neutrino mass bounds applicable to all
generations. Second, there is the question of radiative corrections.
If one estimates the lepton number violating cross-sections naively,
assuming that the appropriate Dirac neutrino mass is of order the
top quark mass, then one finds that the right-handed neutrino Majorana
mass scale must be quite large, greater than $10^{16} \GeV$ at least \harvturn.
This is so large that it is not possible to reconcile it with determinations
of the intermediate scales via gauge coupling unification.
In Ref. \sotenme\ it was pointed out that, when the radiative corrections are
taken into account, this constraint is reduced to a level that is consistent
with gauge coupling unification.
These are the issues which we set out to
understand with the present work. We find that the constraints
are quite severe for the models that we examine; nevertheless these models are
just adequate to the task of producing the observed baryon asymmetry
of the universe
via out of equilibrium $B-L$ violating processes. One can speculate on the
importance of the fact that both gauge-coupling unification and baryogenesis
concerns point to a unique intermediate unification scale.

We also examine a specific non-minimal prescription for fermion
masses. The behaviour of this model is quite different from the
minimal model; as we discuss below, in some sense these two
models encapsulate the  various possibilities.

\newsec{A Minimal Realistic Model}

The minimal Yukawa interactions necessary to generate realistic
fermion masses and mixings at low energies are
\eqn\minyuk{ h_{ij} \psi^i \psi^j H(10) + f_{ij} \psi^i \psi^j
	\sbar{\Delta(126)}.}
The $H(10)$ is actually two real {\bf 10} dimensional representations
of $SO(10)$ combined
into a complex field. The coupling of the $H(10)$ gives the well-known
tree-level mass relations $m_b=m_\tau$, $m_t= m_{\nu_\tau}^{\rm Dirac}$, and
similarly for the other generations. Radiative corrections
upset these relations, making the tau-bottom mass ratio acceptable.
The coupling of the $\Delta(126)$ gives rise to a Majorana mass matrix for the
right-handed neutrinos upon spontaneous breaking of $U(1)_{B-L}$.
A $U(1)_{PQ}$ Peccei-Quinn symmetry \PQ\babmoh\ forbids the appearance
of a Yukawa coupling to the $\sbar{H(10)}$.

The minimal gauge symmetry breaking scheme has $SO(10)$ breaking
through one intermediate stage,
$SO(10)\rightarrow SU(2)_L\times SU(2)_R \times SU(4) \rightarrow
SU(2)_L \times U(1)_Y \times SU(3)_c$. The required scalar representations
are a {\bf 210} or {\bf 54} to break $SO(10)$, a {\bf 126} to break
the intermediate group and generate right-handed neutrino Majorana
masses, and a {\bf 10} to break the electroweak group. As pointed
out in Ref. \moaxion, an extra representation is required
in order to prevent the $U(1)_{PQ}$ symmetry from surviving to the
electroweak scale and thereby creating an unacceptable weak-scale axion.
A {\bf 16} can be used for this purpose.

{\topinsert
	\epsfxsize=5in
	\vbox{\hskip 0.7 true in \epsfbox{cross.eps}}
	\vskip 0.2in
	\ifig\crossfig{  Scalar mixing induced term giving rise to the effective
		{\bf 126}-like coupling of the $H(10)$ at low energies. }
	\epsfxsize=0pt
	\vskip .7cm
\endinsert
}

At first sight, the simplicity of the Yukawa couplings seems to preclude
realistic fermion mixing. However, it has been demonstrated
that mixing in the scalar sector can induce small
corrections which will generate non-trivial, viable mixing and masses for
fermions \babmoh.
Thus the fermion mass model given by eqn. \minyuk\ is very attractive.
There is essentially one free parameter in this model, the $\Delta(126)$
{\it vev}, $v_R$. Given this parameter,
the charged fermion masses and the CKM matrix determine the neutrino masses
and the lepton mixing matrix \babmoh.
The diagram which induces the small admixture of {\bf 126}-like
Yukawa couplings for the $H(10)$ is shown in  \crossfig. The
{\it vevs} which appear are of the (1,3,$\sbar{10}$) content of
the $\Delta(126)$.

The Yukawa couplings for the effective theory just below the $U(1)_{B-L}$
breaking scale can be written in the form
\eqn\pertyuks{ \eqalign{
	\,& \sbar{U_L} (h + f \delta_u) U_R \, \phi_u, \cr
	\,& \sbar{D_L} (h + f \delta_d) D_R \, \phi_d, \cr
	\,& \sbar{N_L} (h - 3f \delta_u) N_R \, \phi_u, \cr
	\,& \sbar{L_L} (h - 3f \delta_d) L_R \, \phi_d, \cr
	\,& \sbar{N_R^c} f N_R \Phi, \cr
	}
}
where $\delta_u$ and $\delta_d$ are (complex) parameters derived from the
scalar mixing diagram illustrated in \crossfig. The dimension-four
Peccei-Quinn invariant operator which induces the mixing shown in
\crossfig\ is
\eqn\mixop{ \Delta(126) \sbar{\Delta(126)} \Delta(126) H(10). }
Therefore the sizes of
$\delta_u$ and $\delta_d$ are determined by the ratio
$v_R^2/M_\Sigma^2$, where $M_\Sigma$ is the mass of the scalar
field $\Sigma(2,2,15) \in \Delta(126)$.
Notice that if $\delta_u = \delta_d$ then quark mixing would vanish.
The difference in $\delta_u$ and $\delta_d$ is generated by $SU(2)_R$
breaking, which splits the degeneracy of the $SU(2)_L$ doublets contained
in $\Sigma$. Phases in $\delta_u$ and $\delta_d$ can be generated by
effective soft Peccei-Quinn breaking terms.

In the following, we will consider a low-energy theory containing
two Higgs doublets. One could consider a low-energy theory
with one Higgs doublet by choosing a linear combination of
$\phi_u$ and $\tilde \phi_d$ to become ultra-massive.
Note that the renormalization group analysis of the Yukawa couplings
becomes complicated in the two-doublet case due to non-linearities
which cannot be ignored. Thus the solutions found in Ref. \babmoh\
cannot be taken over wholly. We are in the process of a full renormalization
group analysis for the Yukawa couplings in this model, and details regarding
the evolution of these couplings are postponed to a future paper
\nextpaper.

There are a few small inconveniences in the determination of the
model parameters from our knowledge of charged fermions. Clearly, the
low-energy parameters can be used only to fix the product $\delta_u f_{ij}$
and the ratio $\delta_u / \delta_d$. Therefore the coupling
matrix $f_{ij}$ is actually determined only to within an overall
scale connected to our inability to determine the scale of
$\delta_u$ and $\delta_d$. This scale
ambiguity translates into an overall scale ambiguity for the
masses of the right-handed neutrinos, and we often write the Majorana
mass matrix for the right-handed neutrinos in the form
\eqn\NRmass{
	M^{Maj}_{ij} = v_R f_{ij} = {1\over R}\, \delta_u f_{ij}\, v_u,}
where $v_u$ is the {\it vev} of $\phi_u$ and $ R= v_u  \delta_u / v_R $.
A further difficulty is connected with the implementation of CP violation.
CP violation is spontaneous, manifested as phases in the propagators
of the $\Sigma$ fields.
and the two independent CP phases can be
rotated into the induced {\it vev} parameters $\delta_u$ and $\delta_d$.
Strictly speaking, because of our lack of knowledge of the phases in the mixing
matrix for right-handed currents, it is not actually possible to
completely determine
the coupling matrices $h_{ij}$ and $f_{ij}$ from low-energy data.
One could work from the GUT scale down, but this shoot and miss
approach is cumbersome. However, for small values of the phases
a perturbative determination is possible, as discussed in Ref. \babmoh;
we determine the couplings
assuming that CP violation is absent, and then calculate the CP-violating
corrections to the fermion masses and mixings. This leaves
only sign ambiguities in
various fermion masses, which can be chosen so that solutions to the
constraints are generated. As commented in Ref. \babmoh, large CP phase
solutions to the constraints are not found.

Once the coupling matrices $h_{ij}$ and $f_{ij}$ and the parameters
$\delta_u$ and $\delta_d$ are prescribed, the
history of the lepton number asymmetry in the
early universe is fixed. The dominant lepton number
violating effect is the decay of the heavy right-handed neutrino.
This out of equilibrium decay, together with C and CP violation, will
create a lepton asymmetry. Subsequently, this lepton asymmetry will
be distributed into a baryon asymmetry by the action of anomalous
$B+L$ violation \leptontrouble\harvturn. Counter to this simple mechanism is
the
effect of the dimension five lepton number violating
operators which are associated with the light neutrino masses
\eqn\dimfive{ {\cal O}_5 = {m_{\nu i}
	 \over v^2} \phi^0 \phi^0 \nu_{Li}\nu_{Li},}
where $v$ is the weak scale {\it vev}, and $\phi^0$ is the neutral
member of a low energy Higgs doublet.

The competition between the CP-violating parts of the right-handed
neutrino decays, which produce a lepton asymmetry, and the
processes $\nu_{Li}\nu_{Li} \rightleftharpoons \phi^0 \phi^0$, which
subsequently destroy it, is governed by the Boltzmann equations
for right-handed neutrino density and the $B-L$ asymmetry. The framework
for this calculation was established in Ref. \markus, where the
necessary rates were calculated and the Boltzmann equations were
derived and integrated for several generic cases. We choose to write the
Boltzmann equations in a slightly different form. Let $Y_i$ indicate
the number density of species $i$, normalized to entropy, and let
\def\YNieq{Y_{N_{Ri}}^{\rm eq}}
\def\YNjeq{Y_{N_{Rj}}^{\rm eq}}
$\Delta_i = Y_{N_{Ri}} - \YNieq$, where $Y^{\rm eq}$
is the equilibrium number density.
Let
$x=M_{N_{R1}}/T$, where $N_{R1}$ is the lightest of the heavy neutrinos,
and let $x_i= m_i/T$ for a generic species, $i$.
The basic decay and scattering processes of interest to us are
\eqn\rcs{ \eqalign{
	D_{ij}&:\;\;\; N_{Ri}\rightarrow \nu_{Lj} \phi^0,\cr
	N_{ij} &:\;\;\;  \nu_{Li}\nu_{Lj}\rightarrow \phi^0\phi^0,\cr
	Hs_{ij} &:\;\;\;  N_{Ri} L_{Lj} \rightarrow Q_L \sbar{t}_R ,\cr
	Ht_{ij} &:\;\;\;  N_{Ri} \sbar{t}_R \rightarrow L_{Lj} Q_L.
	}
}
Then we have the Boltzmann equations
\def\HMN{H(M_{N_{R1}})}
\eqn\BoltDelta{
	\Dx \Delta_i = -\Dx \YNieq - {x\over \HMN} \Delta_i
		\sum_j \left[ {\tilde\gamma}_{Dij}
		+ {{\tilde\gamma}_{Htij} + {\tilde\gamma}_{Hsij} \over \YNieq}
		\right],
}
\eqn\BoltYBL{
     \Dx Y_{(B-L)_i} = {x\over \HMN}
	\sum_j \left[\epsilon_{ij}  \Delta_j{\tilde\gamma}_{Dji}
		- {Y_{(B-L)_i}\over Y^{\rm eq}_{leptons}}
		 S_{ji}^{BL}
	\right],
}
where
\eqn\YBLscatt{
	S_{ji}^{BL} = \left[ {\tilde\gamma}_{Nji}
                + \left(1+{\Delta_j\over\YNjeq}\right) {\tilde\gamma}_{Hsji}
                + {\tilde\gamma}_{Htji} \right],
}
and where $\tilde\gamma_{Dij}$ is the thermally averaged decay rate
for $N_{Ri}\rightarrow \nu_{Lj} \phi^0$,
\eqn\Drate{ \eqalign{
	\tilde\gamma_{Dij} &={ K_1(x_i)\over K_2(x_i) }
		\Gamma(N_{Ri}\rightarrow \nu_{Lj} \phi^0),\cr
  	  &= { K_1(x_i)\over K_2(x_i) } {h_{ji}h_{ji}^*\over 16\pi}M_{N_{Ri}}.
	}
}
The Hubble rate at temperature $T$ has been denoted $H(T)$ above.
The various thermally averaged reduced cross-sections, dimensionalized
to rates by a factor of the entropy density, are given by
\eqn\thrates{
	{\tilde\gamma}_k = {45 T\over 64\pi^6 g_* }
		\int_{\sqrt{s_{min}}/T}^\infty dy\, y^2 K_1(y)
			{\hat\sigma}_k(T^2 y^2)
}
with
\eqn\redcross{ \eqalign{
	{\hat\sigma}_{Htij}(T^2 y^2) &=
		{m_t^2 h_{ji}h_{ji}^*\over {\pi v^2}}
		\left[ 1 - {x_i^2\over y^2}
			+ {x_i^2\over y^2}\ln\left( {y^2-x^2
				+ x_{\phi^0}^2 \over x_{\phi^0}^2}
						\right)
		\right], \cr
	{\hat\sigma}_{Hsij}(T^2 y^2) &=
		{m_t^2  h_{ji}h_{ji}^*\over {2\pi v^2}}
		\left[ {1- {x_i^2\over y^2}} \right]^2, \cr
	{\hat\sigma}_{N_{ij}} &= {1\over 2\pi}\sum_d
		{ h_{id}h_{dj}h_{id}^* h_{dj}^*  }
		{{x_d}^2\over y^2} g_d(y^2/x_d)^2,
	}
}
and $g_*$ is the effective number of degrees of freedom at
$T\simeq 10^{12} \GeV$, $g_*\simeq 120$.
The function $g_i(z)$ is calculated in Ref. \markus, but appears with
a typographical error in that reference. The correct equation is
\eqn\gcorrect{
	g_i(z) = z + {z\over D_i(z)} + {z^2\over D_i(z)^2}
		- \left( 1 + {1+z \over D_i(z)} \right)\ln(1+z),
}
where $D_i(z)$ is defined in Ref. \markus. As in that reference,
we have neglected the interference terms in ${\hat\sigma}_{N_{ij}}$,
assuming that, for a given $i$ and $j$, one heavy neutrino
exchange dominates the others.
The most important of the exhibited processes, of course,
are the heavy neutrino decays and the lepton number violating
scatterings induced by the dimension five operators. The CP violating
parameters $\epsilon_{ij}$ in eqn. \BoltYBL, associated with the
decay rates ${\tilde\gamma}_{Dji}$ are given by
\eqn\epsij{
	\epsilon_{ij} = {1\over \pi h_{ij} h_{ij}^* }
		\sum_{k,l}\, {\rm Im}( h_{kj} h^*_{kl} h_{il}^* h_{ij} )
		\,f(M_{N_{Rl}}^2 / M_{N_{Rj}}^2).
}
Here, $f(z)= \sqrt{z}(1 - (1+z)\ln({{1+z}\over z}))$.
Notice that the quantity
$\epsilon_{ij} \Gamma( N_{R_j} \rightarrow \phi^0 \nu_{L_i} )$,
which appears in the Boltzmann equation for the lepton asymmetries
is not a re-phasing invariant. However, when summed over $i$, as it
appears in the equation for the total lepton asymmetry, it becomes
re-phasing invariant, as it must since the total lepton asymmetry
is independent of the phase convention.

It is also useful to know how $B+L$ is depleted. Given the
rate per unit volume for $B+L$ violating processes as \arnmc\
\eqn\BLrate{
	\gamma_{B+L} = C (\alpha_2 T)^4,
}
it is easy to derive the exponential depletion equation
\Peccei\
\eqn\BLdep{
	Y_{B+L}(x) = Y_{B+L}(0) \, e^{-\kappa x},
}
where $\kappa \simeq C (10^{12} \GeV / M_{N_{R1}})$, $C$ being
${\cal O}(1)$ in the high temperature limit.

{\topinsert
	\epsfxsize=6in
        \vbox{\hskip -1.5 true in \epsfbox{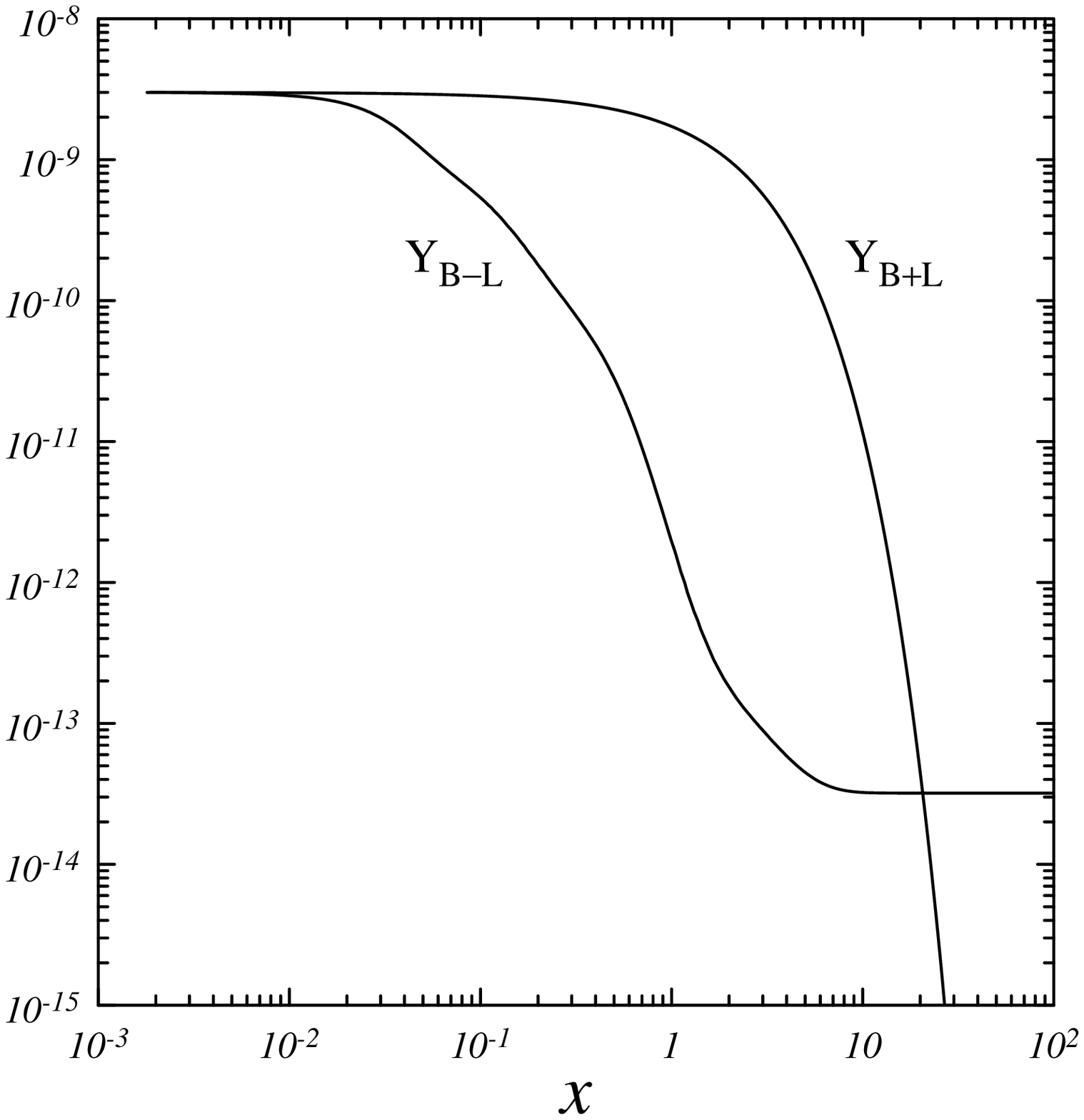}}
	\ifig\depletefig{ Baryon number depletion in the absence of leptonic
	CP violation. The lightest right-handed neutrino has mass
	$M_{N_1} = 2\times 10^{12} \GeV$, and $x=M_{N_1}/T$}
        \epsfxsize=0pt
        \vskip .7cm
\endinsert
}

Before proceeding to the calculation which is relevant for the model at hand,
it is interesting
to calculate the baryon number depletion caused by the $B+L$ and $B-L$
violating processes. We imagine (unrealistically) that the CP violation
in the lepton sector vanishes. By calculating the depletion of an initial
baryon asymmetry as a function of the right-handed scale, we can calculate
the lower bound on that scale, assuming that the initial baryon asymmetry
must survive to low temperatures. This is the standard argument for bounding
neutrino masses, transferred to our specific model and utilizing a full
numerical computation.
We integrate the Boltzmann equations numerically, using
an implicit differencing scheme to insure computational stability
\numrec.
In \depletefig\ we show the depletion of an
initial asymmetry $Y_B \simeq 3\times 10^{-9}$ when the lightest right-handed
neutrino has mass $M_{N_1} = 2\times 10^{12} \GeV$.
The linearity of the equations implies that the initial asymmetry would
be required
to be greater than $10^{-5}$ in order for an acceptable baryon number to
survive
to low temperatures. Alternatively, the mass of the lightest right-handed
neutrino
would be required to be greater than $2\times 10^{13} \GeV$.
This is difficult to accommodate within
the context of this model since the right-handed scale is bounded by gauge
coupling unification criteria. The example shown, with $m_t=150\GeV$,
in the absence of leptonic CP
violation, would be ruled out. This is a generic feature for $SO(10)$ grand
unification; the $U(1)_{I_{3R}}\times U(1)_{B-L}$ breaking scale is bounded
above by the minimum of the $SU(2)_R$ and $SU(4)$ breaking scales, and
therefore must be less than $M_{R_0}^{\rm max} \simeq 3\times 10^{13} \GeV$.
This implies, by a triviality bound discussed in Ref. \sotenme,
that the heaviest right-handed neutrino must be less than approximately
$1.5 M_{R_0}^{\rm max}/g_R \simeq 2\times 10^{14} \GeV$.
There is some freedom in the conclusions, since lowering
$m_t$ will decrease the neutrino Dirac masses and thus diminish the
effect of the lepton-number violating scattering processes.
Again, we defer a catalog of parameter
space until later \nextpaper. Thus we now turn
to leptonic CP violation in order to explain the observed baryon
asymmetry in the context of this model, without recourse to strict
bounds on the top quark mass.

{\topinsert
	\epsfxsize=6in
	\epsfysize=6in
	\vbox{\hskip -1.5 true in \epsfbox{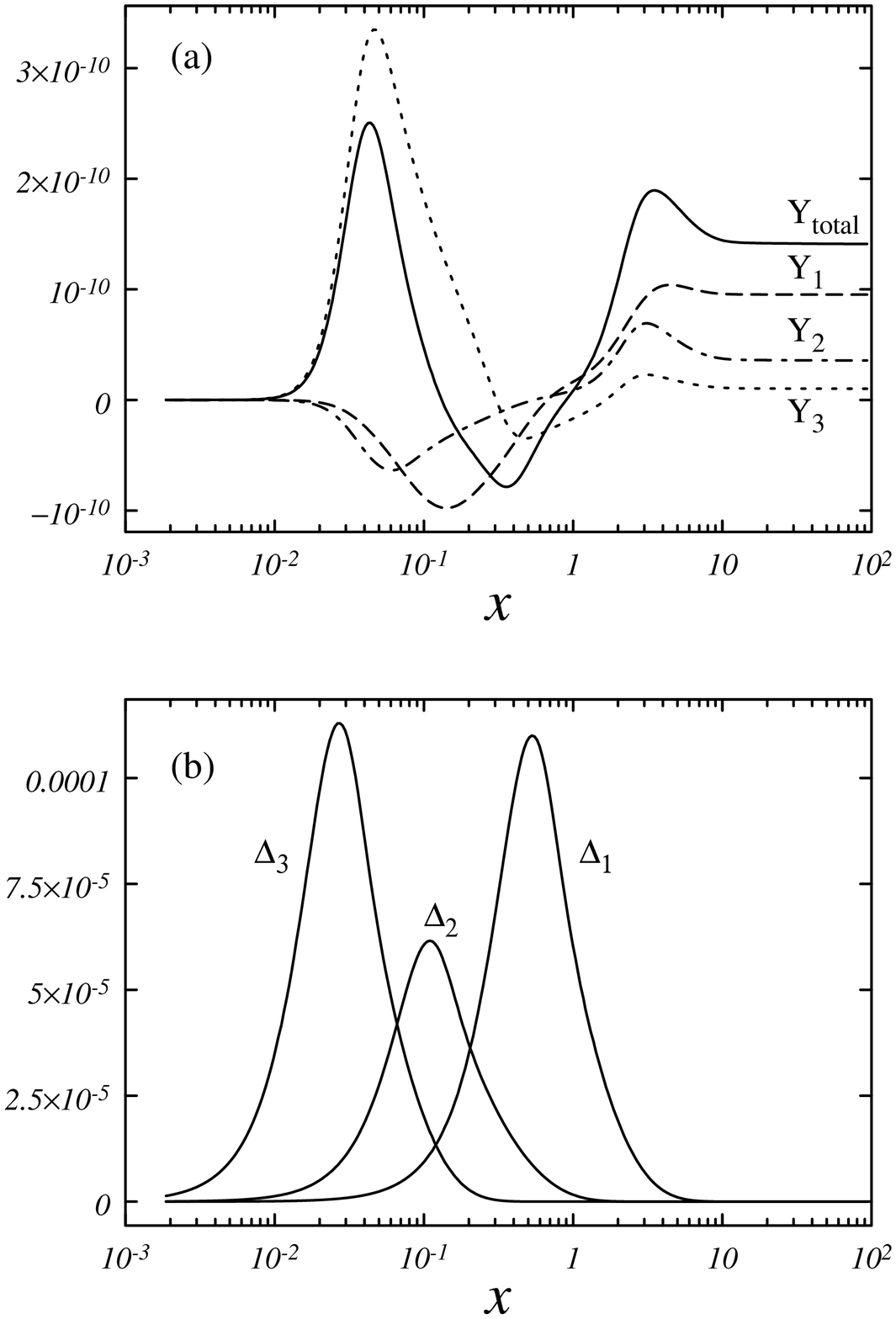}}
	\ifig\bmonefig{ Lepton asymmetry production for Set 1, as described
        in the text, with $1/R=3\times 10^{14}$. See also the Appendix.
	(a) shows the $B-L$ asymmetry for each generation, as well as the
	total asymmetry, normalized to the entropy density.
	(b) shows the $N_R$ deviations from equilibrium, $\Delta_i$, as
	defined in the text.}
	\epsfxsize=0pt
	\epsfysize=0pt
	\vskip .7cm
\endinsert
}

Using the approach of Ref. \babmoh, with appropriate renormalization
group evolution of the Yukawa couplings, we find many solutions for
the Yukawa couplings which give rise to acceptable fermion masses
and mixings. Using these solutions we calculate the CP-violating
parameters $\epsilon_{ij}$, and the neutrino masses and mixings.
Recall that the right-handed neutrino mass scale is an extra input
parameter. There is no special correlation between the coupling
matrices $f_{ij}$ and $h_{ij}$; therefore the leptons are well-mixed,
as indicated by the absence of a large hierarchy in the neutrino
Dirac Yukawa couplings (see the Appendix).
Given these parameters, we calculate the lepton asymmetry.
\bmonefig\ displays the results of one such calculation, with
$1/R= 3\times 10^{14}$ so that the right-handed neutrino masses
are
$M_{N_i} = (2\times 10^{12} \GeV, 6\times 10^{12}\GeV, 3\times 10^{13}\GeV)$.
We have chosen a set of Yukawa couplings
giving as large an asymmetry as we could find. The largest
values of $\epsilon_{ij}$ are of order $10^{-5}$. This is sensible since
the CP-violation in the neutrino sector is related to that in the
quark sector via the unification ansatz. The baryon asymmetry produced
by sphaleron conversion is related to the $B-L$ symmetry by \harvturn
\eqn\htthing{
	\Delta B \simeq {1\over 3} \Delta(B-L).
}
The observed baryon asymmetry is
$Y_B^{\rm obs} \simeq (0.6-1.0)\times 10^{-10}$ \obsB.
The $B-L$ asymmetry
illustrated in \bmonefig\ is sufficient to explain the observed
baryon asymmetry.

Furthermore, we comment that the produced asymmetry is approximately
proportional to the scale of the right-handed neutrino masses. This is
easy to understand, since the decay rates which produce
the $\Delta_i$ are proportional to the masses, and the scattering
processes which destroy $Y_{(B-L)_i}$ are approximately inversely proportional
to the squares of the masses (This is exactly true, of course, at low energies,
but the behaviour at energies near the right-handed neutrino masses is
more complicated). Therefore, we must take the right-handed neutrino masses
as large
as possible \naive. This is the manifestation of the
neutrino mass bounds \leptontrouble\harvturn. Note that we can derive
an upper bound on the right-handed neutrino masses by a ``triviality''
argument, as discussed in Ref. \sotenme; $| f_{ij} |\, \roughly{<}\, 1.5 $,
so that the heaviest right-handed neutrino should satisfy
$M_N \, \roughly{<}\, 1.5 v_R$, implying $1/R \, \roughly{<}\, 3\times
10^{14}$.

In the Appendix, we have collected the specific parameter values for those
sets of Yukawa couplings for which we give results. The results
on $\Delta B$ production are presented
in Table 1.

\vskip 0.5in
\begintable
  | $1/R$ | $N_R$ Masses (GeV) | $\nu_i$ Masses (eV)| $Y_B$  \crthick
  Set 1 | $3\times 10^{14}$ | $\;2\times 10^{12},6\times 10^{12},3\times
10^{13}\;$ | $\;4\times 10^{-6},2\times 10^{-4}, 0.03\;$| $0.9\times
10^{-10}$\cr
  Set 1 | $2\times 10^{14}$ | $\;10^{12},4\times 10^{12},2\times 10^{13}\;$ |
$\;7\times 10^{-6},3\times 10^{-4},0.05\;$| $0.5\times 10^{-10}$\cr
  Set 1 | $5\times 10^{13}$ | $\;3\times 10^{11},10^{12},6\times 10^{12}\;$ |
$\;3\times 10^{-5}, 10^{-3}, 0.2\;$| $6.0\times 10^{-12}$\cr
  Set 2 | $2\times 10^{14}$ | $\;10^{12},4\times 10^{12},2\times 10^{13}\;$ |
$\;7\times 10^{-6},3\times 10^{-4},0.05\;$ | $2.5\times 10^{-12}$\cr
  Set 2 | $5\times 10^{13}$ | $\;3\times 10^{11},10^{12},6\times 10^{12}\;$ |
$\;3\times 10^{-5}, 10^{-3}, 0.2\;$ | $0.8\times 10^{-12}$
\endtable

\vskip 0.1in

\noindent
{\bf Table 1} Some results on baryon asymmetry production and neutrino
masses for the minimal model of fermion masses.
\vskip 0.2in

\newsec{ Non-Minimal Models }

The above class of models illustrates that the out of equilibrium processes
implied by the necessary suppression of $B-L$ violation contain
the hidden potential for production of a baryon asymmetry of the correct
order of magnitude via leptogenesis.
However, the pattern of neutrino masses in these minimal models
is very rigid. We can ask how much of this rigidity is due to the details
of the model. As an attempt to understand this, we investigate certain
non-minimal prescriptions for fermion masses.

The first possible extension is the introduction of extra heavy
{\bf 126} fields. In this way a non-trivial phase structure in the Majorana
mass
matrix can be spontaneously generated. These fields will affect the
Dirac Yukawa couplings of the {\bf 10} exactly as in \crossfig\ so that
mixing and CP-violation in the quark sector will again be suppressed
naturally by ratios of order $v_R^2/M_{\Sigma_i}^2$. However, CP-violation
from the neutrino Majorana mass matrix can be full strength.
Of course, any predictive power is lost with such an addition, unless
we are willing to consider specific ansatze. We content ourselves with
the observation that the net effect of such an addition will be
the introduction of full-strength CP-violating phases which can be rotated
into the neutrino Dirac mass matrix upon diagonalization of the
right-handed neutrino mass matrix. Since the neutrino Dirac Yukawa
couplings are still dominated by the direct couplings to the {\bf 10},
they will not change appreciably in magnitude. Therefore the enhancement
in CP-violation for right-handed neutrino decays will be given
at most by the ratio of the angles which appear. Comparing to the
analysis above, we see that this will provide at most an enhancement
of a factor of five in the produced asymmetry, since the CP phase angles which
appeared above were $\simeq \pi/10$. Thus we do not expect that a qualitatively
different light neutrino mass spectrum is possible with a generic introduction
of extra heavy {\bf 126} fields.

Next, let us examine a particular family of non-minimal ansatze.
Consider the popular Fritzsch type ansatze \frit, with the incorporation
of the Georgi-Jarlskog asymptotic relation $m_s \simeq  m_\mu /3$ \GJ.
A specific model which is phenomenologically interesting is presented
in Ref. \babshaf. We do not address how such a mass matrix ansatz could
arise from the underlying theory. There are several ways in which
non-minimal scalar representations and mixing in the scalar sector
could be introduced in order to generate such mass matrices.
The neutrino Dirac Yukawa couplings take the form
\eqn\bsDy{ h_\nu^{\rm Dirac} = {1\over v_u}
	\left( \matrix{ 0 & \sqrt{m_u m_c} & 0 \cr
			\sqrt{m_u m_c} & 0 & -3\sqrt{m_c m_t} \cr
			0 & -3\sqrt{m_c m_t} & m_t }
	\right),
}
and the Majorana mass matrix for the right-handed neutrinos
takes the form
\eqn\bsMaj{ M_\nu^{\rm Majorana} =
	\left( \matrix{  M_1 e^{i\gamma_1} & 0 & 0 \cr
			0 & M_2 e^{i\gamma_2} & M_3 e^{i\gamma_3} \cr
			0 & M_3 e^{i\gamma_3} & 0 \cr }
	\right).
}
The masses which appear in $h_\nu^{\rm Dirac}$ are, of course, the asymptotic
values obtained by a renormalization group evolution from low energies.
Triviality bounds still require the eigenvalues of $M_\nu^{\rm Majorana}$ to be
bounded by $\simeq 1.5 v_R$, but otherwise we assume the right-handed
neutrino masses are arbitrary.
$M_\nu^{\rm Majorana}$ is diagonalized by the transformation
\eqn\bsMajdiag{
	 U^T M_\nu^{\rm Majorana} U,}
where $U=K V$ is unitary, with $V$ orthogonal and $K$ a diagonal matrix of
phases.
{}From the form of $M_\nu^{\rm Majorana}$ we have
\eqn\Vbs{
	V = \left(\matrix{ 1 & 0 & 0 \cr
			0 & \cos\theta & -\sin\theta \cr
			0 & \sin\theta & \cos\theta
		}
		\right),
}
and we let $\sigma_i$ be the phases in $K$,
\eqn\Kbs{
	K= {\rm diag}( e^{i\sigma_1}, e^{i\sigma_2}, e^{i\sigma_3} ).
}

Because of the imposed symmetries, which have determined the form of the
mass matrices, neutrino masses in this model are very different from the
minimal model considered above. The neutrino states are not ``well-mixed''
at all, and the diagonalization of the right-handed neutrino masses will
not upset the large hierarchy in the Dirac Yukawa couplings. Thus the
right-handed neutrino which has mass $M_1$ will have very weak Dirac Yukawa
couplings, and will decay very slowly. Furthermore, the lepton-number violating
cross-sections for the light family will be highly suppressed.

{\topinsert
	\epsfxsize=6in
	\vbox{\hskip -1.5 true in \epsfbox{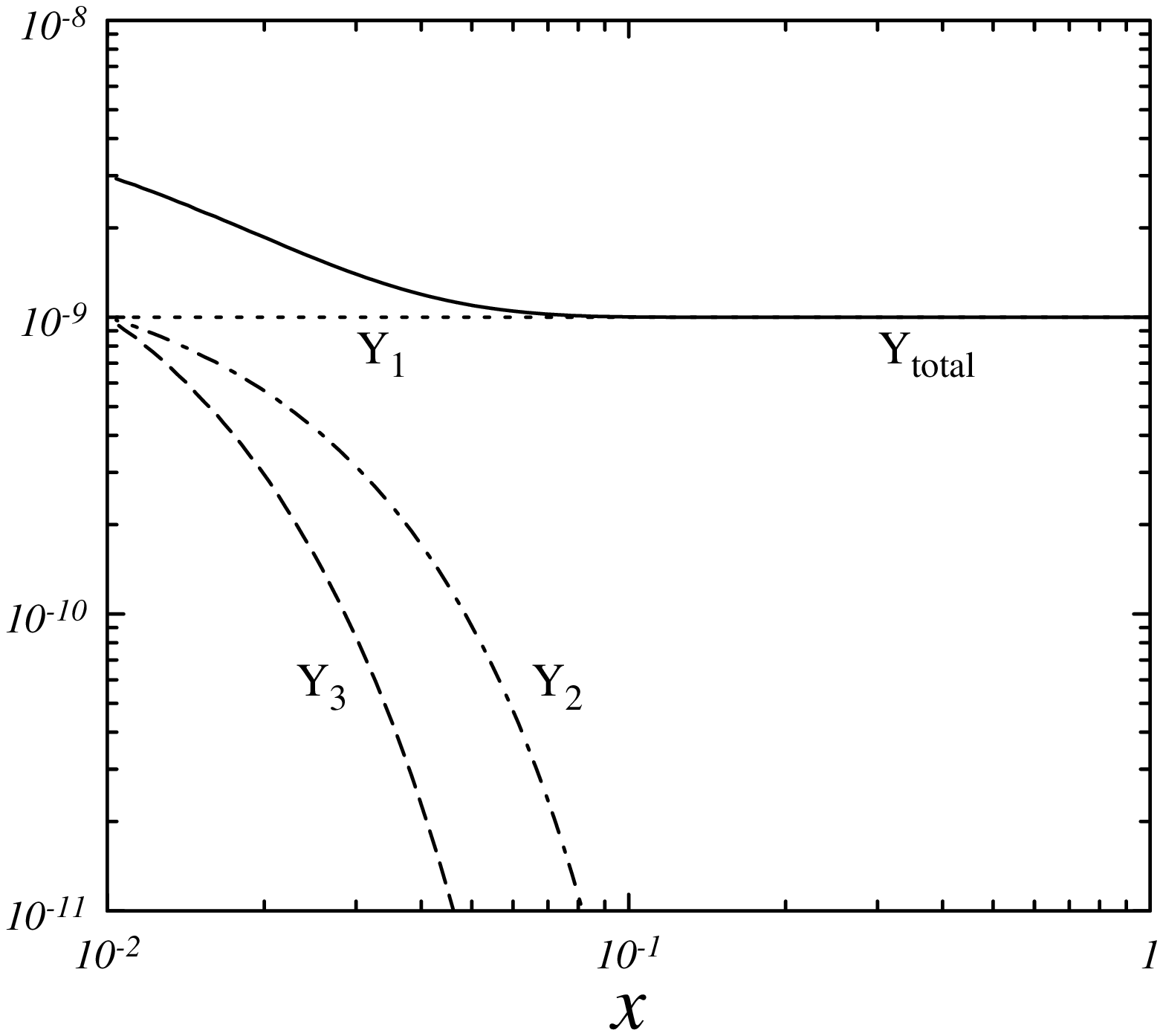}}
	\ifig\depNMfig{ $B-L$ depletion for the Fritzsch ansatz model with
		right-handed neutrino masses all equal $10^{11} \GeV$.
		Notice that the asymmetry associated with the light family
		survives unmolested.}
	\epsfxsize=0pt
	\vskip .7cm
\endinsert
}

The first issue we must examine is whether or not an initial baryon asymmetry
is able to survive the period of lepton-number violating interactions.
As done above for the minimal models, we set the phases in the Majorana mass
matrix to zero and
consider the evolution of an initially nonzero baryon asymmetry from the GUT
scale through the intermediate temperature regime.
The depletion of $B+L$ is complete and is exactly as shown in \depletefig.
In \depNMfig\ we show the depletion of $B-L$, and thus $B$, for right-handed
neutrino masses all equal $10^{11} \GeV$. As we expect from the poor mixing of
the
light generation, approximately one third of the initial asymmetry will survive
the lepton-number violating temperature regime, and thus this model avoids the
neutrino mass bounds. In fact, we find that the right-handed neutrino masses
can be taken as small as $10^9 \GeV$ before lepton-number violating effects
begin
to deplete the asymmetry associated with the light generation. The depletion
factor
$Y_{B-L}^{\rm final} / Y_{B-L}^{\rm initial}$ as a function of the lightest
right-handed
neutrino mass is given in Table 2.

\vskip 0.5in
\begintable
 $M_{N1}$ (\GeV) | $Y_{B-L}^{\rm final} / Y_{B-L}^{\rm initial}$ \crthick
   $10^{10}$    |    1.0 \cr
   $10^{9} $    |    0.8 \cr
   $10^{8} $    |    0.1
\endtable
\vskip 0.1in

\noindent
{\bf Table 2} Depletion of $B-L$ asymmetry as a function of the lightest
right-handed neutrino mass for the Fritzsch ansatz model.
\vskip 0.2in

Given the above, we might expect that little more could be said about this
model.
However, the introduction of nonzero phases in the Majorana mass matrix
creates a constraint of a different type. Because the decay rate of the first
generation right-handed
neutrino is highly suppressed, it remains out of equilibrium for much
longer than it otherwise would. As the temperature decreases, the
lepton-number violating scattering
processes quietly become unimportant, but the decays of this neutrino continue
to produce a $B-L$ asymmetry. This state of affairs can continue over two
decades of temperature. The resulting asymmetry can be very large, and this
must
be avoided.
The results of some calculations (far from exhaustive) for this model are shown
in Table 3. We have chosen to parametrize the right-handed neutrino
Majorana matrix in terms of its positive eigenvalues, $\mu_i$, and the phases
$\sigma_i$.
In terms of the mass matrix parameters, $\mu_1 = M_1$,
$\mu_{2,3} = {1\over 2}| M_2 \pm \sqrt{ M_2^2 + 4 M_3^2} | $.
The mixing angle $\theta$ is determined from the $\mu_i$,
$\tan\theta=\sqrt{\mu_3/\mu_2}$.
The phases $\sigma_i$ were chosen arbitrarily. Note that the entropy production
associated
with the decay of the light $N_R$ is no longer negligible, due to its great
deviation from equilibrium;
using the estimate $\Delta s \simeq M_N/g_* T_{\rm decay}$ \entropyprod\markus,
the decrease of the asymmetry is
approximately a factor of two, and this is incorporated in the results.

\vskip 0.5in

\begintable
   $ \mu_1,\mu_2,\mu_3\, (\GeV)$ | $ \sigma_1\; , \sigma_2\; ,\sigma_3\; $ |
$\nu_i$ Masses (eV)| $Y_B$  \crthick
   $10^{12},10^{12},10^{12}$ | $20^\circ, 45^\circ, 60^\circ$ |$\;
10^{-11},6\times 10^{-3}, 5\;$| $2\times 10^{-7}$\cr
   $10^{11},10^{11},10^{11}$ | $20^\circ, 45^\circ, 60^\circ$
|$\;10^{-10},0.06,50$ | $10^{-8}\;$\cr
   $\;3\times 10^{10},3\times 10^{11},3\times 10^{11}\;$| $20^\circ, 45^\circ,
60^\circ$ |$\;3\times 10^{-10},0.02,16\;$|$3\times 10^{-9} $\cr
   $\;3\times 10^{10},3\times 10^{11},3\times 10^{11}\;$| $6^\circ, 5^\circ,
4^\circ$ |$\;3\times 10^{-10},0.02,16\;$|$2\times 10^{-10} $\cr
   $\;5\times 10^{10},5\times 10^{10},5\times 10^{10}\;$| $20^\circ, 45^\circ,
60^\circ$ |$\;2\times 10^{-10},0.1,97\;$|$3\times 10^{-10} $\cr
   $10^{10},10^{10},10^{10}$ | $20^\circ, 45^\circ, 60^\circ$ |
$10^{-9},0.6,490$ | $ 2\times 10^{-14}$
\endtable
\vskip 0.1in

\noindent
{\bf Table 3} Production of a baryon asymmetry via leptogenesis in the Fritzsch
 ansatz model.

\vskip 0.2in

\newsec{Discussion}

The two models examined above have contrasting cosmological consequences,
and there is a sense in which they display the two natural possibilities.
The minimal model posseses no special symmetries in the fermion couplings to
scalars, and therefore displays ``democratic'' behaviour in the language
of Ref. \markus. On the other hand, the symmetries which determine the form
of the right-handed Majorana mass matrix in the non-minimal model will not
allow a relaxation of the Dirac hierarchy via mixing, and this model
is ``correlated'' in the language of Ref. \markus. To display a behaviour
intermediate to this would almost certainly require the introduction
of small parameters which allow the  Majorana mass
matrix to interpolate between the symmetric form of the non-minimal model and
a more generic form. It is not clear how such small parameters could be
introduced
in a natural way.

We have seen that the minimal models probably do not allow an initial baryon
asymmetry
to survive the intermediate temperature period at the required level. However,
the intermediate scales are such that CP violation in the lepton sector
can produce a viable $B-L$ asymmetry; sphaleron re-processing will
create a baryon asymmetry of the observed size.
The survival question for an initial baryon asymmetry thus becomes
irrelevant in this model.
The
consequences for neutrino mass phenomenology are, of course, somewhat grim.
Certainly the light neutrino masses are well below the level relevant to
MSW solar neutrino oscillations \solaranal. They are also an order
of magnitude too large for a vacuum oscillation solution \solaranal.
It is interesting that our bounds on the
$U(1)_{B-L}\times U(1)_{I_{3R}}$ breaking scale
drive us  to a unique gauge symmetry breaking chain, the minimal one,
$SO(10)\rightarrow SU(2)_L\times SU(2)_R \times SU(4)$.

The results for the non-minimal model which we examined were in marked
contrast to those of the minimal model. Due to the imposed symmetric
mass matrix structure, an initial baryon asymmetry is able to survive the
intermediate temperature period for all interesting values of the right-handed
neutrino masses. However, the introduction of leptonic CP
violation is problematic. Because of the large suppression of
the decay rate for one species of right-handed neutrino, the production
of a $B-L$ asymmetry can escalate out of control. This effect can be countered
by choosing a slight hierarchy in the right-handed neutrino masses
and by choosing the angles $\sigma_i$ of order $\pi/10$. We could find no
solution with larger phases, and this seems a little uncomfortable,
since there is no mechanism which could ensure this situation.
These choices give rise to light neutrino masses which can be
phenomenologically
interesting, as indicated by the one viable entry in Table 3, with
light neutrino masses $m_{\nu i} \simeq 3\times 10^{-10}\eV, 0.02\eV, 16\eV$.
Notice that the third generation neutrino naturally comes out quite heavy,
approaching the upper bound from cosmological overclosure arguments
\cowsik, making it a hot dark matter candidate.
This is another interesting coincidence of the model.

We comment that the identification of the measured baryon asymmetry with
a GUT-scale generated asymmetry, as is possible in the non-minimal
model, may be plagued with other difficulties. For example, an initial baryon
asymmetry will be destroyed by an inflationary period which is probably
required to dilute the population of magnetic monopoles which are
generated at the GUT scale \preskill. There may also be a problem
with monopoles generated at the intermediate scale, though there
exists a mechanism for the annihilation of these monopoles \monann.
Obviously such issues depend on details other than the fermion
mass spectrum, and it would be beyond the scope of this paper to
address them here.

G.J. would like to acknowledge a discussion with K.S.~Babu regarding
Ref. \babmoh.
G.J. was supported by DOE grant DEFG02-90-ER 40560, and
T.G. was supported by NSF contract PHY-91-23780.

\vskip 0.5in
{\bf Appendix }
\vskip 0.2in

In this appendix we list the two sets of parameter values in the minimal model
for which
we present results. Most of this does not require comment. We note
only that we can accommodate a phenomenologically acceptable value
for $|V_{cb}|$, and that the ratio $m_d/m_u$ is typically about 1.5, which
is slightly smaller than the value which appears in Ref. \gassleut, but this
does not worry us excessively. The neutrino Dirac Yukawa coupling matrix
which we give is expressed in the basis where the right-handed neutrino
Majorana mass matrix has been diagonalized.
$J$ is the re-phasing invariant measure of CP violation in the
quark sector \jarl. Notice that the CP violation
in the quark sector
is controlled by the difference in  $\arg\,\delta_u$ and $\arg\,\delta_d$;
in Set 2 we have lowered $\arg\,\delta_u$, but the value of $J$ has increased.

\vskip 0.4in
\noindent Set 1:
$$ \matrix{
	m_t=150 \GeV & m_c=1.22 \GeV & m_b= -4.35 \GeV \cr
	\tan\beta = -47.1 & |V_{cb}| = 0.045 & |V_{ub}|= 0.006 \cr
	m_u = 4 \MeV & m_d= 6.5 \MeV & m_s = 135 \MeV \cr
	J \simeq 2\times 10^{-6} & \arg\,\delta_u = 9^\circ & \arg\,\delta_d =
13^\circ \cr
	}
$$

$$
	M_{N_R} = {1\over R} ( 0.006 \GeV, 0.022 \GeV, 0.12 \GeV)
$$

$$
	h_\nu^{\rm Dirac} = \left(
		\matrix{ 0.016 & 0.045 & -0.066 \cr
			 0.045 & 0.13 & -0.19 \cr
			 -0.066 & -0.19 & 0.29 \cr}
			\right)
$$

$$
	\epsilon_{ij} = \left(
	\matrix{ 2.7 \times 10^{-6}  & -7.6\times 10^{-6} & -2.2\times 10^{-6} \cr
		 1.4 \times 10^{-6} & -2.3\times 10^{-7} & -6.8\times 10^{-7} \cr
		 4.7 \times 10^{-7} & -1.5\times 10^{-6} & 2.9\times 10^{-6} \cr
	}
	\right)
$$

\vskip 0.5in
\noindent Set 2:

As above except $\arg\,\delta_u = 4^\circ$, $J\simeq 10^{-5}$, $m_u= 4.6 \MeV$,
and

$$
	\epsilon_{ij} = \left(
	\matrix{ 9.7\times 10^{-7} & -7.5\times 10^{-6} & 1.6\times 10^{-5} \cr
	     -5.3 \times 10^{-7} & -6.9\times 10^{-6} & 1.6\times 10^{-5} \cr
		-2.3\times 10^{-7} & -6.3\times 10^{-6} & 1.4\times 10^{-5} \cr
	 }
        \right)
$$

\listrefs     
\bye